\documentclass{ws-ijmpb}

\begin{document}

\markboth{J. M. Morbec and K. Capelle}
{Exact and approximate relations for the spin-dependence of the
exchange energy in high magnetic fields}

%
\catchline{}{}{}{}{}
%

\title{Exact and approximate relations for the spin-dependence of the
exchange energy in high magnetic fields}

\author{J. M. Morbec\footnote{Communicating author. Email: jmorbec@yahoo.com.br} \ and K. Capelle}

\address{Departamento de F\'{\i}sica e Inform\'atica\\ Instituto de F\'{\i}sica de S\~ao Carlos\\ 
Universidade de S\~ao Paulo\\
Caixa Postal 369, 13560-970 S\~ao Carlos, SP, Brazil}



\maketitle


\begin{abstract}
The exchange energy of an arbitrary collinear-spin many-body system in 
an external magnetic
field is a functional of the spin-resolved charge and current densities,
$E_x[n_{\uparrow},n_{\downarrow},{\bf j}_{\uparrow},{\bf j}_{\downarrow}]$.
Within the framework of density-functional theory (DFT), we show that the
dependence of this functional on the four densities can be fully reconstructed
from either of two extreme limits: a fully polarized system or a completely
unpolarized system. Reconstruction from the limit of an unpolarized system
yields a generalization of the Oliver-Perdew spin scaling relations from
spin-DFT to current-DFT. Reconstruction from the limit of a fully polarized
system is used to derive the high-field form of the local-spin-density
approximation to current-DFT and to magnetic-field DFT.
\end{abstract}


\section{Introduction}

The exchange energy is a fundamental quantity of many-body physics. While it 
can be calculated analytically and exactly for simple quantum fluids, such 
as the three-dimensional interacting electron gas in the absence of magnetic 
fields, its evaluation for realistic, spatially inhomogeneous, many-electron 
systems is possible only numerically, {\em e.g.} by means of density-functional
theory (DFT) within the local-spin-density approximation (LSDA).\cite{vbh72,gl76} In collinear spin-DFT (SDFT), the exact exchange energy is a functional of 
the spin-resolved charge density $E_x[n_\uparrow,n_\downarrow]$. In strong 
external or internal magnetic fields, orbital magnetism must be accounted for 
on the same footing as spin magnetism, and SDFT is replaced by current-DFT 
(CDFT)\cite{vr1,vr2} or magnetic-field DFT (BDFT).\cite{bdft1,bdft2}

In the present paper we discuss exact and approximate expressions for the
spin-dependence of the exchange energy $E_x$. We show that, for arbitrary 
collinear-spin many-body systems, the dependence of $E_x$ on spin-resolved
charge and current densities can be fully reconstructed from either of two 
extreme limits: a fully polarized system or a completely unpolarized system. 
Reconstruction from the limit of an unpolarized system yields a generalization 
of the Oliver-Perdew spin scaling relations\cite{op79} from SDFT to CDFT.
The alternative possibility of reconstruction from the limit of a fully 
polarized system results in explict expressions for the high-field exchange 
energy, which can be used as input for the LSDA of CDFT or of BDFT.

\section{Exact properties: spin scaling}

Since exchange acts only between like-spin electrons, the exchange energy 
of an arbitrary many-electron system (homogeneous or not, in a magnetic field 
or not) can be written as the sum of two contributions, one arising entirely 
from the spin up subsystem, the other from the spin down subsystem
\begin{equation}
E_x= \tilde{E}_x^{\uparrow} + \tilde{E}_x^{\downarrow}.
\label{exdecomp}
\end{equation}
This decomposition is valid as long as there is a preferred direction relative
to which one can define spin up and spin down, {\em i.e.} for collinear
magnetism. It only ceases to be valid in noncollinear situations, where the 
quantization axis changes as a function of space. The exchange energy then 
acquires a spin off-diagonal contribution which, in principle, can be treated 
on the same footing as the diagonal term.\cite{sdwdft} Here, we are only 
concerned with collinear situations, occuring in unidirectional magnetic 
fields. Under these circumstances, Eq.~(\ref{exdecomp}) always holds. 

According to the general theorems of CDFT,\cite{vr1,vr2} the exchange energy 
of an arbitrary (collinear) many-body system is a functional of the 
spin-resolved charge and current densities $n_{\uparrow}({\bf r})$, 
$n_{\downarrow}({\bf r})$,  ${\bf j}_{\uparrow}({\bf r})$ and 
${\bf j}_{\downarrow}({\bf r})$, where ${\bf j}({\bf r})={\bf j}_{\uparrow}({\bf r})+{\bf j}_{\downarrow}({\bf r})$ is the paramagnetic current density, 
related to the gauge invariant physical current density by 
\begin{equation}
{\bf j}_{\rm phys}({\bf r}) = {\bf j}({\bf r}) + {e\over{mc}} 
n({\bf r}){\bf A}({\bf r}).
\label{jdef}
\end{equation}
By symmetry,
$E_x[n_{\uparrow},n_{\downarrow},{\bf j}_{\uparrow},{\bf j}_{\downarrow}] 
= E_x[n_{\downarrow},n_{\uparrow},{\bf j}_{\downarrow},{\bf j}_{\uparrow}]$.
The decomposition (\ref{exdecomp}) then implies
\begin{equation}
E_x[n_{\uparrow},n_{\downarrow},{\bf j}_{\uparrow},{\bf j}_{\downarrow}]=
\tilde{E_x}[n_{\uparrow},{\bf j}_{\uparrow}] + \tilde{E}_x[n_{\downarrow},{\bf j}_{\downarrow}],
\label{exfunc}
\end{equation}
where the spin up and spin down contribution are given by the same 
functional, although their values for specific densities, 
$\tilde{E}_x^{\uparrow}$ and $\tilde{E}_x^{\downarrow}$, are, generally, 
different.

We now define two related functionals, the exchange energy of a spin 
unpolarized system\cite{footnote1}
\begin{equation}
E_x^{NP}[n,{\bf j}]:=
E_x\left[\frac{n}{2},\frac{n}{2},\frac{\bf j}{2},\frac{\bf j}{2}\right]
\label{exnp}
\end{equation}
and that of a fully polarized system
\begin{equation}
E_x^{FP}[n,{\bf j}]:= E_x[n,0,{\bf j},0].
\label{exfp} 
\end{equation}
By evaluating Eq.~(\ref{exfunc}) at the densities of an unpolarized 
system, we obtain
\begin{equation}
\tilde{E}_x[n,{\bf j}]=\frac{1}{2}E_x^{NP}[2n,2{\bf j}],
\end{equation}
from which it follows that we can write $E_x$ as
\begin{equation}
E_x[n_{\uparrow},n_{\downarrow},{\bf j}_{\uparrow},{\bf j}_{\downarrow}]=
\frac{1}{2}E_x^{NP}[2n_{\uparrow},2{\bf j}_{\uparrow}] +
\frac{1}{2}E_x^{NP}[2n_{\downarrow},2{\bf j}_{\downarrow}],
\label{exfuncnp}
\end{equation}
which holds for arbitrary polarizations. By evaluating this at the 
densities of a fully polarized system, we find $\tilde{E}_x[n,{\bf j}]
=E_x^{FP}[n,{\bf j}] $, which implies that the general exchange energy can 
also be written as
\begin{equation}
E_x[n_{\uparrow},n_{\downarrow},{\bf j}_{\uparrow},{\bf j}_{\downarrow}]=
E_x^{FP}[n_{\uparrow},{\bf j}_{\uparrow}] + E_x^{FP}[n_{\downarrow},{\bf j}_{\downarrow}].
\label{exfuncfp}
\end{equation}
The two limiting cases of the general functional 
$E_x[n_{\uparrow},n_{\downarrow},{\bf j}_{\uparrow},{\bf j}_{\downarrow}]$ are 
thus related by
\begin{equation}
E_x^{FP}[n,{\bf j}]={1\over 2} E_x^{NP}[2n,2{\bf j}].
\label{limitrel}
\end{equation}
Equations (\ref{exfuncnp}) and (\ref{exfuncfp}) show that the exchange 
energy has the remarkable property that its functional form at arbitrary 
polarization is completely determined by either that at zero polarization 
or that at full polarization, without requiring interpolation between 
both limits or additional calculations inbetween them.

If its current dependence is ignored, Eq.~(\ref{exfuncnp}) reduces to the 
spin-scaling relations of Oliver and Perdew,\cite{op79} which are frequently 
used in spin-density-functional theory (SDFT) to connect spin-density 
functionals for arbitrarily polarized systems to density functionals obtained
from many-body calculations for an unpolarized system.\cite{spinscaling1,spinscaling2,spinscaling3} 
Equation~(\ref{exfuncnp}) is the generalization of this spin scaling from SDFT 
to CDFT, and may be employed in the same way.

Equations reconstructing the general functional from the fully polarized
limit, such as Eq.~(\ref{exfuncfp}), are less used in DFT, where one usually
constructs a spin-dependent functional by starting from the unpolarized
situation. However, they are useful in the special case of the high-field
limit, where analytical results for the fully polarized exchange energy of 
the homogeneous interacting electron gas are available for the first 
($L=0$)\cite{dg71} and second ($L=1$)\cite{mc08} Landau level.

\section{Approximate expressions: high-field exchange energy of the 
homogeneous electron gas}

In the homogeneous electron gas in a constant unidirectional magnetic field
$B$, the charge density and the physical current density are spatially 
constant. For sufficiently high magnetic fields only the lowest Landau level 
contributes, and the electrons are fully spin polarized. The per-volume 
exchange energy $e^{FP}_x(n,B)=E^{FP}_x(n,B)/V$ in this situation has been 
evaluated analytically by Danz and Glasser.\cite{dg71} In the limit in which 
all electrons occupy the bottom of the lowest Landau level, $L=0$, they find, 
\begin{equation}
e_x^{FP}(n,B)=2\pi e^2 \lambda^2 n^2 \left[\ln \left(
\lambda^3 n\right)+2.11779 \right], 
\label{dgfp}
\end{equation}
where $\lambda=\left(\hbar c/eB\right)^{1/2}$ is the magnetic length. 

Expressions applicable to still higher fields, as well as generalization to 
weaker fields in which the restriction to the bottom of the lowest Landau level 
is removed (while maintaining that to the lowest Landau level, $L=0$, itself) 
can be 
found in the original reference.\cite{dg71} In a recent paper\cite{mc08} we 
reevaluated the exchange energy in these limits, corrected a few minor problems
in the original equations,\cite{dg71,bcr74} and discussed the range of magnetic
fields and densities where each of them applies. 

For still weaker magnetic fields, two complications can take place: orbital
magnetism now aquires a contribution of higher Landau levels, and spin 
magnetism is reduced, as the system is not necessarily fully polarized.
We dealt with the former complication by deriving an explicit expression 
for $L=1$.\cite{mc08} Here, we consider partial and vanishing spin 
polarization, which may occur for lower fields, in particular when
the electron $g$ factor is smaller than $2$. 

In a first step, we note that the many-body calculations\cite{dg71,mc08,bcr74}
yield $e_x^{FP}(n,{\bf B})$ as function of the charge density $n$ and the 
magnetic field 
$B$, {\em i.e.} the variables used in BDFT, but not of the spin and current 
densities used in CDFT. The transition from one set of variables to the other 
is affected by recalling that in the homogeneous electron gas in a constant 
magnetic field ${\bf j}_{\rm phys}({\bf r}) \equiv 0$, so that, from 
Eq.~(\ref{jdef}), ${\bf B}(n,{\bf j})=-\frac{mc}{e} \nabla \times ({\bf j}/n)$. 
From Eq.~(\ref{exfuncfp}) we thus have\cite{footnote3}
\begin{eqnarray}
e_x(n_{\uparrow},n_{\downarrow},{\bf j}_{\uparrow},{\bf j}_{\downarrow})=
\nonumber \\
e_x^{FP}\left(n_{\uparrow},{\bf B}= -\frac{mc}{e}\nabla\times
\left({{\bf j}_{\uparrow}\over n_{\uparrow}}\right)\right)
+ e_x^{FP}\left(n_{\downarrow},{\bf B}=-\frac{mc}{e}\nabla\times
\left({{\bf j}_{\downarrow} \over n_{\downarrow}}\right)\right).
\label{vortscal}
\end{eqnarray}

Equation (\ref{vortscal}) provides the exchange energy for arbitrarily 
polarized systems, as a function of the spin-resolved charge and current 
densities. As an example, we use it to generate from Eq.~(\ref{dgfp}) the 
exchange energy of an unpolarized system. This can be done either by using 
(\ref{vortscal}) to generate the result for an arbitrarily polarized system
and subsequently substitute the densities of an unpolarized system, or, 
equivalently, directly from the special case of Eq.~(\ref{limitrel}). 
The result is
\begin{equation}
e_x^{NP}(n,{\bf B})=\pi e^2 \lambda^2 n^2\left[\ln \left(\lambda^3n\right)+1.42464\right].
\label{dgexgenNP}
\end{equation}

All other relations, valid for different magnetic field ranges,\cite{dg71,mc08}
can be treated in the same way. We have chosen to exemplify the procedure for 
Eq.~(\ref{dgfp}), because this allows an independent numerical test of our 
previous assertion (which was based on explicit calculation\cite{mc08}) that 
the Danz-Glasser expression (\ref{dgfp}) is correct, and not the alternative 
result\cite{bcr74}
\begin{equation}
e^{FP,alt}_x(n,{\bf B})
=2\pi e^2 \lambda^2 n^2 \left[\ln \left(\lambda^3 n\right) 
+ 2.32918 \right],
\label{bcrfp}
\end{equation}
which, when scaled to the unpolarized case, becomes
\begin{equation}
e^{NP,alt}_x(n,{\bf B})= \pi e^2\lambda^2 n^2\left[\ln\left(\lambda^3 n\right)+1.63603\right].
\label{bcrnp}
\end{equation}
In the unpolarized case, Takada and Goto\cite{takadagoto} numerically 
calculated $e_x$, so that we can use their data to check the validity of 
Eqs.~(\ref{dgexgenNP}) and (\ref{bcrnp}), and thus also of (\ref{dgfp})
and (\ref{bcrfp}). We find that (\ref{dgexgenNP}) is in perfect agreement with 
numerical data extracted from figure 2(b) of Takada and Goto,\cite{takadagoto} 
whereas (\ref{bcrnp}) deviates from the numerical data by $\sim 10\%$ in the 
metallic density range.

\section{Conclusions}

Two types of spin-scaling relations have been demonstrated for the 
current-dependent exchange energy, one employing the fully polarized limit 
and the other the unpolarized limit. Simple connections, valid for arbitrary 
many-body systems, exist between these limits and the general functional 
$E_x[n_{\uparrow},n_{\downarrow},{\bf j}_{\uparrow},{\bf j}_{\downarrow}]$.
In the special case of the homogeneous electron gas, such connections, applied 
to many-body calculations in the fully polarized limit,\cite{dg71,mc08,bcr74} 
yield explicit formulas that allow to use available numerical data for 
unpolarized systems\cite{takadagoto} to arbitrate between conflicting expressions at
full polarization. Results for arbitrary polarization can be used as input for 
the local-spin-density approximation to current-density-functional theory.

\section*{Acknowledgements}
This work received financial support from FAPESP and CNPq.


\section*{References}
\begin{thebibliography}{0}
\bibitem{vbh72} U.~von~Barth and L.~Hedin, {\it J. Phys. C} {\bf 5}, 1629 
(1972).
\bibitem{gl76} O.~Gunnarsson and B. I.~Lundqvist, {\it Phys. Rev. B} {\bf 13}, 
4274 (1976).
\bibitem{vr1} G.~Vignale and  M.~Rasolt, {\it Phys. Rev. Lett.} {\bf 59}, 2360 
(1987).
\bibitem{vr2} G.~Vignale and M.~Rasolt, {\it Phys. Rev. B} {\bf 37}, 10685 
(1988).
\bibitem{bdft1} C. J. Grayce and R. A. Harris, {\it Phys. Rev. A} {\bf 50}, 
3089 (1994).
\bibitem{bdft2} F. R.~Salsbury and R.~A.~Harris, {\it J. Chem. Phys.} {\bf 107}, 
7350 (1997).
\bibitem{op79} G. L. Oliver and J. P. Perdew, {\it Phys. Rev. A} {\bf 20}, 397 
(1979).
\bibitem{sdwdft} K.~Capelle and L.~N.~Oliveira, {\it Phys. Rev. B} {\bf 61}, 
15228 (2000).
\bibitem{footnote1} NP standing for ``not polarized'' --- the more natural 
choice of UP (unpolarized) may lead to confusion with the spin up contribution.
\bibitem{spinscaling1} J. P. Perdew, A. Ruzsinszky, J. Tao, V. N. Staroverov, 
G. E. Scuseria, and G. I. Csonka, {\it J. Chem. Phys.} {\bf 123}, 062201 (2005).
\bibitem{spinscaling2}  J. P. Perdew, K. Burke and M. Ernzerhof,
{\it Phys. Rev. Lett.} {\bf 77}, 3865 (1996).
\bibitem{spinscaling3} K. Capelle and G. Vignale,
{\it Phys. Rev. Lett.} {\bf 86}, 5546 (2001).
\bibitem{dg71} R. W. Danz and M. L. Glasser, {\it Phys. Rev. B} {\bf 4}, 94 
(1971).
\bibitem{mc08} J. M. Morbec and K. Capelle, {\it Phys. Rev. B} {\bf 78}, 085107 (2008).
\bibitem{bcr74} B. Banerjee, D. H. Constantinescu and P. Rehak,
{\it Phys. Rev. D} {\bf 10}, 2384 (1974).
\bibitem{footnote3} Note that we do not distinguish functionals of the current 
from those of the magnetic field by a new symbol, since they can always be 
distinguished through their arguments.
\bibitem{takadagoto} Y. Takada and H. Goto, {\it J. Phys. Condens. Matter} 
{\bf 10}, 11315 (1998).
\end{thebibliography}
\end{document}